\documentclass[copyright,creativecommons,noderivs,noncommercial]{eptcs}
\providecommand{\event}{}
\providecommand{\volume}{??}
\providecommand{\anno}{20??}
\providecommand{\firstpage}{17}
\providecommand{\eid}{??}

\usepackage{latexsym}
\usepackage{amssymb}
\usepackage{epsfig}
\usepackage{graphics}
\usepackage{url}
\usepackage{color}

\newcommand{\sys}[1]{\langle #1 \rangle}
\newcommand{\set}[1]{\{ #1 \}}
\newcommand{\List}[1]{\sys{#1}}

\newcommand\State{S}
\newcommand\Init{I}
\newcommand\Obs{\mathcal{O}}
\newcommand\Act{ACT}
\newcommand\Prop{Prop}
\newcommand\prot{P}
\newcommand\Runs{\mathcal{R}}
\newcommand\knows{K}
\newcommand\view{\mathit{view}}
\newcommand\sends[1]{\stackrel{#1}{\Longrightarrow}}
\newcommand\fm{\mathit{F}_m} 
\newcommand\comp{\phi^{c}}

\newcommand\Nat{\mathbb{N}}
\newcommand\power{\mathcal{P}}

\newtheorem{definition}{\bf Definition}[section]

\newtheorem{proposition}[definition]{\bf Proposition}

\title{How to Work with Honest but Curious Judges?\\(Preliminary Report)}

\author{
Jun Pang
\institute{University of Luxembourg\\
Faculty of Sciences, Technology and Communication \\
6, rue Richard Coudenhove-Kalergi, L-1359 Luxembourg}
\email{jun.pang@uni.lu}
\and
Chenyi Zhang
\institute{University of Luxembourg\\
Faculty of Sciences, Technology and Communication \\
6, rue Richard Coudenhove-Kalergi, L-1359 Luxembourg}
\email{chenyi.zhang@uni.lu}
}

\def\titlerunning{How to Work with Honest but Curious Judges?}
\def\authorrunning{J. Pang \& C. Zhang}

\begin{document}
\maketitle

\begin{abstract}

The three-judges protocol, recently advocated by Mclver and Morgan
as an example of stepwise refinement of security protocols,
studies how to securely compute the majority function to reach a final verdict
without revealing each individual judge's decision.
We extend their protocol in two different ways for an arbitrary number of $2n+1$ judges.
The first generalisation is inherently centralised,
in the sense that it requires a judge as a leader who collects information from others,
computes the majority function, and announces the final result.
A different approach can be obtained by slightly modifying the
well-known dining cryptographers protocol,
however it reveals the number of votes rather than the final verdict.
We define a notion of conditional anonymity in order to analyse these two solutions.
Both of them have been checked in the model checker MCMAS.

\end{abstract}

\section{Introduction}
\label{sec:intro}

With the growth and commercialisation of the Internet,
users become more and more concerned about their anonymity and privacy in the digital world.
Anonymity is the property of keeping secret the identity of the user who has performed a certain action.
The need for anonymity arises in a variety of situations,
from anonymous communications, electronic voting, and donations to postings on electronic forums.

Anonymity (untraceability) was first proposed by Chaum~\cite{chaum_88}
in his famous dinning cryptographers protocol (DCP).
After that, a great deal of research has been carried out on this topic
and various formal definitions and frameworks for analysing anonymity have been developed in the literature.
For example, Schneider and Sidiropoulos analysed anonymity with CSP~\cite{SS96}.
They used substitution and observable equivalence to define anonymity in CSP.
In their framework, the automatic tool FDR~\cite{Low96} was used to check the equivalence of two processes.
Kremer and Ryan~\cite{KR05} analysed the FOO92 voting protocol with the applied pi calculus
and proved that it satisfies anonymity partially with an automated tool ProVerif~\cite{Bla01}.
Chothia {\it et al.}~\cite{COPT07} proposed a general framework based on
the process algebraic verification tool $\mu$CRL~\cite{BCLOP+07} for checking anonymity.
Anonymity can be captured in a more straightforward way in epistemic logics,
in terms of agents' knowledge,
and model checkers for epistemic logics, such as MCK~\cite{meyden_04},
LYS~\cite{EO07} and MCMAS~\cite{LQR09}, have been applied to DCP.
Other works, including~\cite{HO03,HS04,BP05,DPP07,CPP07}, have considered probabilistic anonymity.

In all aforementioned works, DCP has been taken as a running example.
DCP is a method of anonymous communication, in that 
it allows for any member of a group to multicast data to other members of the group,
meanwhile it guarantees sender anonymity.
In DCP, all participants first set up pairwise shared secrets using secret channels,
then each participant announces a one-bit message.
If a participant does not want to send a message,
the one-bit message is the XOR of all shared one-bit secrets that he owns.
Otherwise, he announces the opposite.
In order to achieve \emph{unconditional anonymity}, the protocol requires secret channels,
which is difficult to achieve in practice.
Despite its simplicity and elegance,
DCP has been criticised for its efficiency and its vulnerability to malicious attacks.
Several methods, such as~\cite{waidner_89,GJ04}, have been proposed to fix these problems,
but they all make the protocol much more complex.
Notably, Hao and Zieli{\'n}ski~\cite{hao_09} recently presented anonymous veto networks to solve DCP efficiently,
which only requires two rounds of broadcast.
While the original solution of Chaum~\cite{chaum_88} is unconditionally secure,
the solutions proposed in~\cite{waidner_89,GJ04,hao_09} are computationally secure,
as their security is based on the assumption of the intractability of some well-known NP problems.

In essence, DCP implements a secure computation of the boolean OR function from all participants
(which inherently assumes that at most one participant holds the boolean value $1$),
while the individual input bit is kept privacy. 
A general problem is to compute the fuction $F(x_1, x_2, \ldots, x_n)$
without revealing any individual's $x_i$.
The three-judges protocol, recently advocated by Mclver and Morgan~\cite{MM09}
as an example of stepwise refinement of security protocols, 
computes the majority function $F(x_1,x_2,x_3)$ out of three booleans $x_i$ with $i\in\set{1,2,3}$ 
to reach a final verdict without revealing each individual judge's decision $x_i$.\footnote{
The example is taken from a talk by Mclver and Morgan, entitled
``Sheherazade’s Tale of the Three Judges:
An example of stepwise development of security protocols".}
Mclver and Morgan's protocol relies on the 1-out-of-2 oblivious transfer by Rivest~\cite{rivest_99}
and all communications in the protocol are public.
At the end of the protocol all three judges know the majority verdict,
but no one knows more than their own judgement.
More details about this protocol can be found in Section~\ref{sec:problem}.

In our point of view,
the three-judges protocol can be regarded as another standard example for 
formal definition and analysis of anonymity.
Unlike DCP, the three-judges protocol securely computes
a majority function rather than the boolean OR function,
which actually gives rise to some difficulties when
we generalise Mclver and Morgan's solution for an arbitrary number of $2n+1$ judges
in Section~\ref{sec:centralised}.
Our first generalisation is centralised,
in the sense that it requires one judge as a leader of the group.
The leader collects information from others,
computes the majority function, and then announces the final result.
As the leader plays a quite distinct role from the other judges in the protocol,
in certain situations he may know more than necessary due to
the asymmetric design of the protocol.
A second solution is obtained by slightly modifying the dining cryptographers protocol
which is thus inherently symmetric (see Section~\ref{sec:symmetric}).
However, this solution reveals the number of votes (for `guilty') rather than the final verdict.
Therefore, both of the two presented solutions in the paper are imperfect,
anonymity for judges are \emph{conditional} in the sense that
in certain scenarios their decisions are allowed to be deduced.
A formalisation of \emph{conditional anonymity} in a temporal epistemic logic
is given in Section~\ref{sec:anonymity},
which is based on a formal description of the \emph{interpreted system} model~\cite{fagin_95}.
In Section~\ref{sec:analysis}, both solutions are modelled and checked in MCMAS~\cite{LQR09},
a model checker for verification of multi-agent systems.
In the end, we discuss other possible (computational) solutions 
and conclude the paper with future works in Section~\ref{sec:conclusion}.

\section{Description of The Three-Judges Protocol}
\label{sec:problem}

In this section we present the three-judges protocol due to Mclver and Morgan~\cite{MM09}.
Three \emph{honest but curious} judges communicate over the internet to reach a verdict by majority, and the final verdict
is `guilty' if and only if there are at least two judges holding a decision `guilty'.\footnote{
An honest judge follows the protocol strictly,
but he is also curious to find out the other judges' decisions.}
However, once the verdict is announced, each judge is allowed to deduce no more information than his own decision 
as well as the published verdict. To this point, we write $J_i$ with 
$i\in\set{0,1,2}$ for judge $i$, with $d_i$ taking value from $\set{0,1}$ for $J_i$'s private decision, where `$1$' denotes
`guilty' and `$0$' denotes `innocent'. We write $\fm^{n}:\set{0,1}^n\rightarrow\set{0,1}$ for the 
majority function out of $n$ boolean variables. 

One may find that the anonymity security requirement for the three-judges protocol is not as straightforward
as what is in the (three) dining cryptographers protocol (DCP). For instance, it is not necessarily the case
that $J_1$'s decision is always kept secret to $J_2$, typically if $d_2=0$ and $\fm^{3}(d_1,d_2,d_3)=1$.

\subsection{Oblivious Transfer}
McIver-Morgan's solution to the three-judges problem applies Rivest's 1-out-of-2 oblvious transfer protocol (OT)~\cite{rivest_99}.
OT guarantees unconditional security, but it needs private channels and a `trusted initialiser'.
We briefly describe the protocol as follows. The scenario has three parties 
Alice ($A$), Bob ($B$) and a Trusted Initialiser ($T$), where Alice owns messages
$m_0$, $m_1$, and Bob will obtain $m_c$ with $c\in\set{0,1}$ from Alice in a way that
the value $c$ remains secret. It is assumed that there are private channels established between 
$A$, $B$ and $T$, the operator `$\oplus$' is `exclusive or', and messages $m_0,m_1$
are bit strings of length $k$, i.e., $m_0,m_1\in\set{0,1}^k$.
The protocol proceeds as below.
\begin{enumerate}
\item $T\sends{r_0,r_1}A$ and $T\sends{d,r_d}B$, with $r_0,r_1\in\set{0,1}^k$ and $d\in\set{0,1}$.
\item $B\sends{e}A$ with $e=c\oplus d$,
\item $A\sends{f_0,f_1}B$ where $f_0=m_0\oplus r_e$, $f_1=m_1\oplus r_{1-e}$.
\end{enumerate}
In the end, it is verifiable that Bob is able to compute $m_c=f_c\oplus r_d$.\footnote{
The essential idea of this protocol is that after $T$ generates two keys $r_0$
and $r_1$, both keys are sent to $A$ and only one key is sent to $B$ in a randomized way.
Then $B$ can let $A$ encrypt her messages in the `correct' message-key combination such that
$B$ can successfully retrieve $m_c$ after $A$ sends both encrypted messages to $B$.
Since the key sent to $B$ is chosen by $T$, $A$ has no way to deduce which message is
actually decrypted by $B$, and since $B$ obtains only one key
he knows nothing about the other message, and since $T$ quits after the first step
he knows nothing about both messages.} 
The extended version for 1-out-of-n oblivious transfer based on this protocol is straightforward~\cite{rivest_99}. 

\subsection{The McIver-Morgan's Protocol}
\label{ssec:mm-solution}

A solution to solve three-judges protocol has been proposed by McIver and Morgan~\cite{MM09}.
We rephrase their protocol in this section. For notational convenience, we replace judges $J_1$, $J_2$ and $J_3$
by $A$, $B$ and $C$ respectively, with their decisions $a$, $b$ and $c$. 1-out-of-2 Oblivious transfer (OT)
is treated as a primitive operation, so that $A\sends{x/y}B$ means $A$ sends
either $x$ or $y$ to $B$ in the way of OT (i.e., $B$ is to choose a value out of $x$ and $y$). 
The protocol can be presented as follows,
where `$\oplus$' is `exclusive or', `$:=$' denotes variable definition, and `$\equiv$' denotes logical equivalence.

\begin{enumerate}
\item $B$ generates $b_\wedge$ and $b_\wedge'$ satisfying $b_\wedge\oplus b_\wedge'\equiv b$.
\item $B$ $\sends{b_\wedge/b_\wedge'}$ $C$ by OT, then $c_\wedge :=
\left\{
\begin{array}{l}
    b_\wedge' \hspace{5pt}\mbox{ if }\hspace{5pt} c=1,\\
    b_\wedge \hspace{5pt}\mbox{ if }\hspace{5pt} c=0.
  \end{array}
\right. $
\item $B$ generates $b_\vee$ and $b_\vee'$ satisfying $b_\vee\oplus b_\vee'\equiv\neg b$.
\item $B$ $\sends{b_\vee/b_\vee'}$ $C$ by OT, then $c_\vee :=
\left\{
\begin{array}{l}
    b_\vee \hspace{5pt}\mbox{ if }\hspace{5pt} c=1,\\
    b_\vee' \hspace{5pt}\mbox{ if }\hspace{5pt} c=0.
  \end{array}
\right. $
\item $B$ $\sends{b_\wedge/b_\vee}$ $A$ by OT, and $C$ $\sends{c_\wedge/c_\vee}$ $A$ by OT,
then $A$ announces: $\left\{
\begin{array}{l}
    b_\wedge\oplus c_\wedge \quad \hspace{10pt}\mbox{ if }\hspace{5pt} a=1,\\
    \neg(b_\vee\oplus c_\vee) \hspace{5pt}\mbox{ if }\hspace{5pt} a=0.
  \end{array}
\right. $
\end{enumerate}

In this protocol, $A$ only needs to know the result $b\wedge c$ (i.e., whether the other judges
have both voted for `guilty') if he is holding a decision `innocent', or $b\vee c$ (i.e., whether 
at least one of the other judges has voted for `guilty') if he is holding `guilty'.
The rest of the protocol is focused on how to generate two bits $b_\wedge$ and $c_\wedge$ 
satisfying $b_\wedge\oplus c_\wedge = b\wedge c$, and two bits $b_\vee$ and $c_\vee$ 
satisfying $b_\vee\oplus c_\vee = \neg(b\vee c)$, in a way that the individual values of $b$ and $c$ are hidden. 
Both constructions rely on the primitive operation OT.
In the case of $b\wedge c$, first we know that the value of $b_\wedge$ and $b_\wedge'$ are both independent of 
$b$. If $c=1$ then $(b\wedge c)\equiv b$, therefore $C$ needs to get $b_\wedge'$ to ensure
$b_\wedge\wedge c_\wedge\equiv b$. If $c=0$ then we have $(b\wedge c)\equiv 0$, for which $C$ ensures 
$b_\wedge\wedge c_\wedge\equiv 0$ by letting $c_\wedge=b_\wedge$.\footnote{
However, revealing both $b_\wedge$ and $b_\wedge'$ will let $C$ uniquely determine the value of $b$.}
Oblivious transfer ensures that $B$ does not know $c$ since whether $b_\wedge$ or $b_\wedge'$ being
transferred to $C$ is up to the value of $c$. The construction of $\neg(b\vee c)$ can be done in a similar way.

The anonymity requirement for this protocol depends on the actual observations of each judge.
Superficially, if a judge's decision differs from the final verdict, then he is able to deduce that
both other judges are holding a decision different from his. Therefore, we may informally state anonymity
for the case of three judges as that each judge is not allowed to know the other judge's decisions provided
that the final verdict coincides his own decision. We will present a generalised definition of anonymity
requirement for $2n+1$ judges in Section~\ref{sec:anonymity}.

\section{A Generalisation of McIver-Morgan's Solution}
\label{sec:centralised}

As described in McIver-Morgan's solution for three judges,
judge $A$ can be regarded as in the leading role of the whole protocol,
who collects either $b\wedge c$ or $b\vee c$ based on his own decision.
(To be precise, $A$ picks up $b\wedge c$ if $a=0$, or $b\vee c$ if $a=1$.)
Based on this observation, it is therefore conceivable to
have a protocol in which one judge takes the lead, and the other judges only need to send 
their decisions to the leader in an anonymous way. However, it
is not quite clear so far to us that this pattern yields a satisfactorily
anonymous protocol when there are more than three judges.
In this section, we present a protocol
which guarantees only a limited degree of anonymity.

Suppose we have judges $J_0, J_1, \ldots, J_{2n}$ with
their decisions $d_0, d_1, \ldots, d_{2n}$. Without loss of generality, we
let $J_0$ be the leader. We then group the rest $2n$ judges
into $n$ pairs, for example, in the way of $(J_1,J_2), (J_3,J_4),\ldots, (J_{2n-1},J_{2n})$.
Now the similar procedure described in McIver-Morgan's solution
can be used to generate $d_{2i-1}\wedge d_{2i}$ and
$d_{2i-1}\vee d_{2i}$ for all $1\leq i\leq n$.\footnote{
To be more precise, each $J_{2i-1}$ first generates $d^\wedge_{2i-1}$ and
$d^\vee_{2i-1}$, then $J_{2i}$ generates $d^\wedge_{2i}$ and
$d^\vee_{2i}$ based on $d_{2i}$ and using OT, so that $d_{2i-1}\wedge d_{2i}=d^\wedge_{2i-1}\oplus d^\wedge_{2i}$
and  $d_{2i-1}\vee d_{2i}=d^\vee_{2i-1}\oplus d^\vee_{2i}$.}
We first illustrate our solution in the case of five judges.
Suppose judge $J_0$'s decision $d_0$ is $1$, then $J_0$ needs to know for the other four
judges if at least two are holding `guilty'.
Superficially, he may poll one of the two formulas
$d_1\wedge d_2$ and $d_3\wedge d_4$. If one of the two
is true then he knows the verdict is $1$ (for `guilty'). However, both
formulas are only sufficient but not necessary for the final verdict to be true.
The formula equivalent to the statement ``whether at least two judges
are holding `guilty''' is
$\varphi^{\frac{2}{4}}=(d_1\wedge d_2)\vee(d_3\wedge d_4)\vee((d_1\vee d_2)\wedge(d_3\wedge d_4))$,
where we use  $\varphi^{\frac{x}{y}}$ for a boolean formula on decisions of $y$ judges with
at least $x$ judges having their decision~$1$.
Similarly, if $d_0$ is $0$, $J_0$ needs to know whether there
are at least three out of four judges deciding on `guilty', which
can be stated as $(d_1\wedge d_2\wedge d_3)\vee(d_1\wedge d_2\wedge d_4)\vee(d_1\wedge d_3\wedge d_4)\vee(d_2\wedge d_3\wedge d_4)$. 
After a simple translation, we get an equivalent formula
$\varphi^{\frac{3}{4}}=(d_1\vee d_2)\wedge(d_3\vee d_4)\wedge((d_1\wedge d_2)\vee(d_3\vee d_4))$,
which is just  $\varphi^{\frac{2}{4}}$ with all the conjunction operators `$\wedge$'s flipped to `$\vee$'s, 
and all the disjunction operators `$\vee$'s flipped to `$\wedge$'s. 
Overall we have the following proposition.

\begin{figure}[ht]
\begin{center}
\hspace{0.5cm}\input{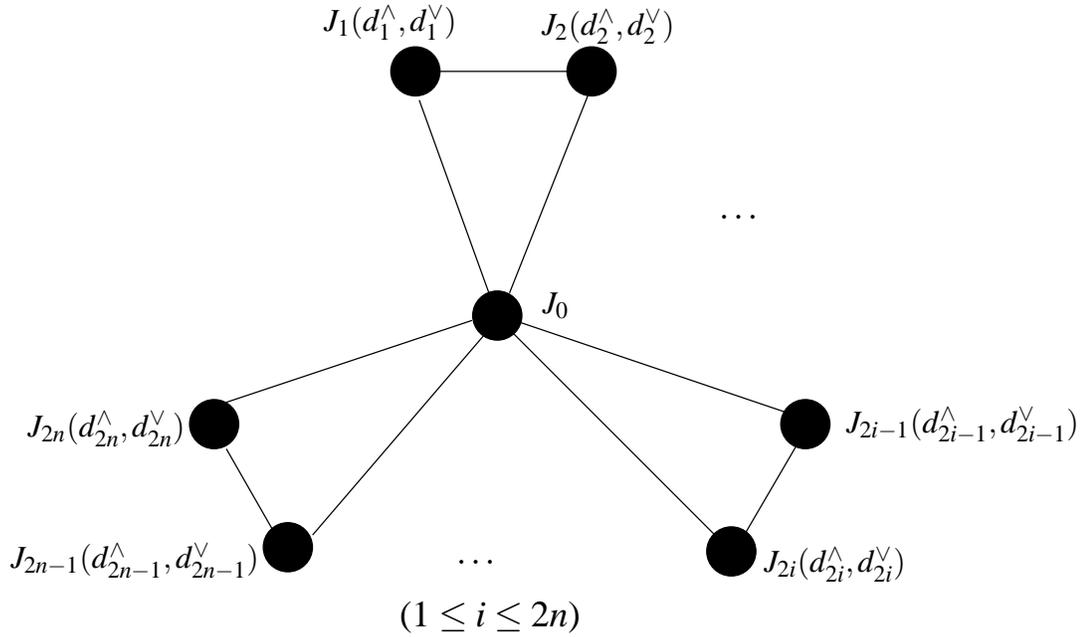}
\end{center}
\caption{A generalisation of McIver-Morgan's judges protocol.}
\label{fig:morgan-judges}
\end{figure}

\begin{proposition}
The formulas $\varphi^{\frac{n+1}{2n}}$ and $\varphi^{\frac{n}{2n}}$ can be constructed by a finite 
number of conjunctions and/or disjunctions from the set of formulas 
$\set{d_{2i-1}\wedge d_{2i}}_{1\leq i\leq n}\cup\set{d_{2i-1}\vee d_{2i}}_{1\leq i\leq n}$.
\end{proposition}

Intuitively, since revealing both $d_{2i-1}\wedge d_{2i}$
and $d_{2i-1}\vee d_{2i}$ gives $J_0$ the actual number of judges in $\set{J_{2i-1},J_{2i}}$
who has voted for `guilty' (plainly, a value in $\set{0,1,2}$),  he will have enough
information to deduce the exact number of judges who has voted `guilty' out of $2n$ in total.
The general protocol for $2n+1$ judges is illustrated in Figure~\ref{fig:morgan-judges},
the lines between judges indicate communications.
An undesirable consequence of this generalisation is that for each pair of judges $J_{2i-1}$ and $J_{2i}$,
if $d_{2i-1}=d_{2i}$ then $d_{2i-1}$ and $d_{2i}$ are both revealed to $J_0$.

\section{A DCP-Based Solution}
\label{sec:symmetric}

In this section, we describe a symmetric solution for computing the majority function based on DCP.\footnote{
This extension to DCP seems to already exist in the literature.
The description closest to ours can be found in~\cite{waidner_89}.
Caroll Morgan also suggests this solution independently from us.}
In DCP, three or more cryptographers sitting in a circle cooperate to make sure
that the occurrence of a certain action, i.e. sending a message, is made known to everyone,
while the cryptographer who has actually performed the action remains anonymous.
They achieve this goal by executing an algorithm which involves coin toss. 
Each neighbouring pair of cryptographers generates a shared bit, by flipping a coin;
then each cryptographer computes the XOR of the two bits shared with the neighbours,
then announces the result -- or the opposite result, if that cryptographer wants to perform the action.
The XOR of the publicly announced results indicates whether such an action has been made.
In the end no individual cryptographer knows who has reported the opposite result.

\begin{figure}[ht]
\begin{center}
\hspace{-1.5cm}\input{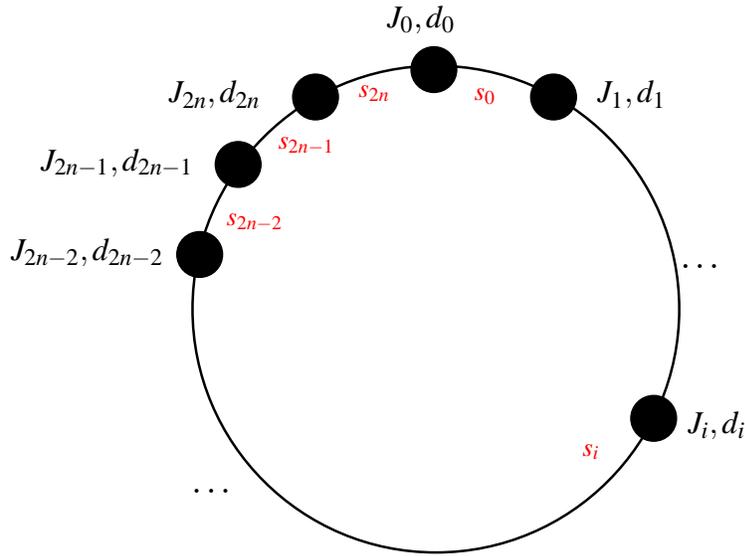}
\end{center}
\caption{A DCP based solution to the judges protocol.}
\label{fig:dcp-judges}
\end{figure}

To extend the DCP technique for $2n+1$ judges protocol,
we require each neighbouring pair of judges $(J_i,J_{i+1})$ with $i\in\{0,\ldots,2n+1\}$\footnote{
Implicitly, the indices are taken modulo $2n+2$ (or a number bigger than $2n+1$).}
in a ring (see Figure~\ref{fig:dcp-judges})
shares one secret $s_i\in\{0,1,\ldots,2n+1\}$.
$s_i$ is used with sign ``+" by $J_i$, with sign ``-" by $J_{i+1}$.
Each judge adds his decision $d_i\in\{0,1\}$ and the sum of his two secrets ($s_{i-1}$ and $s_i$)
with the appropriate signs, and announces the result $(s_i-s_{i-1}+d_i)_{\mid 2n+2}$.
All judges then add up the announced numbers (modulo $2n+2$).
It is easy to see that each secret $s_i$ has been added and subtracted exactly once,
the final sum is just the number of judges who have voted for `guilty', i.e.
$(\sum^{2n}_{i=0}(s_i-s_{i-1}+d_i)_{\mid 2n+2})_{\mid 2n+2}=\sum^{2n}_{i=0}d_i$.
Unlike the solution in the previous section
where only the majority of decisions is made public,
the number of votes are known to every judge
in this symmetric solution where no central leader is needed.
This gives rise to possible attacks, for instance,
the coalition of a group of judges might find out the decisions made by the rest of judges,
if the final sum corresponds to the sum of their votes.

\section{Formalising Anonymity}
\label{sec:anonymity}

Sometimes functionality and anonymity are seemingly contradicting requirements.
For example, in the case of three judges, if one judge discovers that his
decision is different from the final verdict, he will immediately know that both
other judges have cast a vote that is different from his in the current run.
This is why anonymity requirement needs to be specified
\emph{conditional} to the result of each particular run. In other words,
an anonymity specification must be made consistent to what a judge is
legally allowed to know.\footnote{
We are aware of existing works on measuring information leakage,
most of them are developed in a probabilistic setting.
Instead, we aim to formalise a notion of conditional anonymity within an epistemic framework.}

Since all the judges are honest, they make their decisions before a protocol starts. 
Let $\Runs(\prot)$ be the set of runs generated by a protocol $\prot$, and for each $r\in\Runs(\prot)$, 
we write $r(d_i)$ for $J_i$'s decision and $r(v)$ for the final verdict in $r$. 
Anonymity can be defined in terms of compatibility. 
For example, the anonymity property for protocol $\prot$ of \emph{three} judges can  be stated as 
for all judges $J_i$ and $J_j$ with $i,j\in\{0,1,2\}$, $i\neq j$ and $d_i\neq v$,
and for all runs $r\in\Runs(\prot)$, there exists
$r'\in \Runs(\prot)$ such that $r(d_j)\neq r'(d_j)$, $r(d_i)=r'(d_i)$,
$r(v)=r'(v)$, and that $i$ cannot tell the difference between $r$ and $r'$.
Intuitively, this means that given a run $r$ for $i$, if $r(d_j)$ is \emph{compatible}
with both $i$'s decision and the final verdict, then the negation of $r(d_j)$
must also be compatible with $i$'s observation over $r$.
We will show how to generalise this definition by means of
temporal and epistemic logic based specifications.
In addition, a protocol $\prot$ for computing majority of $n$ judges is said to be \emph{functionally correct} 
if in the end of $r$ we have $r(v)=\fm^n(r(d_1),r(d_2),\dots,r(d_n))$ for all runs $r\in\Runs(\prot)$.

\subsection{Interpreted System Model}

To this point we present a formal description of the underlying model
which follows the standard \emph{interpreted system} framework of Fagin \textit{et al.}~\cite{fagin_95},
where there is a finite set of agents $1,2,\dots, n$ and a finite
set of atomic formulas $\Prop$.
The execution of a protocol is modelled as a finite transition system
$\sys{\State, \Init, \Act, \set{\Obs_i}_{1,\dots, n}, \pi, \tau}$, where \begin{itemize}
\item $\State$ is a finite set of states,
\item $\Init\subseteq\State$ is a set of initial states,
\item $\Act$ is a finite set of joint actions,
\item $\Obs_i$ is the observation function of agent $i$, such that $\Obs_i(s)$ is
      the $i$ observable part of a state $s\in\State$,
\item $\pi:\State\rightarrow \power(\Prop)$ is an interpretation function,
\item $\tau:\State\times\Act\rightarrow\power(\State)$ is an evolution (or transition) function.
\end{itemize}
A global state is a cartesian product of the local states of
the agents as well as that of the environment, i.e., $\State=\State_1\times\State_2\times\dots\times\State_n\times\State_e$.
Similarly, we have $\Init=\Init_1\times\dots\times\Init_n\times\Init_e$, and
$\Act = \Act_1\times\Act_2\times\dots\times\Act_n\times\Act_e$. 
We write $s(i)$ and $a(i)$ for the $i$-th part of a state $s$ and an action $a$, respectively.
An agent $i$ is allowed to observe his local state as well as part of the environment state.
The local protocol $\prot_i$ for agent $i$ is a function of type $\State_i\rightarrow\power(\Act_i)$,
mapping local states of agent $i$ to sets of performable actions in $\Act_i$. A protocol $\prot_e$ for
the environment $e$ is of type $\State_e\rightarrow\power(\Act_e)$.
A run $r$ is a sequence $s_0s_1s_2\dots$, satisfying that 
there exists $a\in\Act$ such that $a_m(i)\in\prot_i(s_m(i))$ and
$s_{m}\in\tau(s_{m-1},a)$ for all agents $i$ and $m\in\Nat$. Each transition requires simultaneous inputs
from all agents in the system, and during each transition system time is updated by
one. For each run $r$, $(r, m)$ denotes the $m$-{th} state in $r$ where $m\in\Nat$.
In general, a protocol $P$ is a collection of all agents local protocols, i.e.,
$\prot=\set{P_i}_{i\in\set{1,\dots,n}}\cup\set{P_e}$, and 
$\Runs(\prot)$ is the set of generated runs when the agents execute their local protocols together with $P_e$.

Anonymity requirements can be defined by temporal and epistemic logic in an
interpreted system as generated from a protocol. 
The formulas are defined in the following language,
where each propositional formula $p\in\Prop$ and $i$ denotes an agent.\footnote{
This language can be regarded as a sub-logic used in the model checker MCMAS~\cite{LQR09}.}

\[\phi,\psi := p\mid \neg\phi\mid \phi\wedge\psi\mid \knows_i\phi\mid EX\phi\mid EG\phi\mid E(\phi\ U\psi)\]

The epistemic accessibility relation $\sim_i$ for agent $i$ is defined as
$s\sim_it$ iff $\Obs_i(s)=\Obs_i(t)$. We do not define group knowledge, distributed knowledge and common knowledge
in this paper since in this case study knowledge modality $\knows$ suffices our purpose, and also because
the judges are honest and they do not collude to cheat (e.g., by combining their knowledge).
The temporal fragment of the language follows  
standard CTL (computational tree logic)~\cite{Eme90}, and the semantics of the our formulas are presented as follows. 

\begin{itemize}
\item $s\models p$ iff $p\in\pi(s)$,
\item $s\models \neg\phi$ iff $s\not\models \phi$,
\item $s\models \phi\wedge\psi$ iff $s\models\phi$ and $s\models\psi$,
\item $s\models \knows_i\phi$ iff $s'\models\phi$ for all $s'$ with $s\sim_is'$,
\item $s\models EX\phi$ iff there exists a run $r$ such that $(r,0)=s$ and $(r,1)\models\phi$,
\item $s\models EG\phi$ iff there exists a run $r$ such that $(r,0)=s$ and $(r, m)\models\phi$ for all $m\in\Nat$,
\item $s\models E(\phi\ U\psi)$ iff there exists a run $r$ such that $(r,0)=s$, and there is $m\in\Nat$ satisfying
      $(r,m)\models\psi$ and $(r,m')\models\phi$ for all $0\leq m'< m$.
\end{itemize}

The other standard CTL modalities not appearing in our syntax include $AX$, $EF$, $AF$, $AG$ and $AU$, 
as they are all expressible by the existing temporal modalities. Similar argument holds for the propositional 
logic connectives $\vee$ and $\Rightarrow$. For example, we have $EF\phi$ iff $E(true\ U\phi)$,
$AF\phi$ iff $\neg EG \neg\phi$, and $AG\phi$ iff $\neg EF\neg\phi$, where $true\equiv\neg(p\wedge\neg p)$
for some $p\in\Prop$.

\subsection{Conditional Anonymity}
\label{ssec:conditionalanonymity}

We assume that each judge $J_i$ makes his decision $d_i$ at the beginning of a protocol execution, 
and $v\in\set{1, 0, \bot}$ denotes the final verdict indicating whether there are at least half of the judges
have voted for `guilty', in particular $v=\bot$ denotes that the final verdict is yet to be announced.
The functionality of the protocol can also be verified by checking if every judge eventually
knows the final verdict as $\fm^{2n+1}(d_0,d_2,\dots,d_{2n})$ for $2n+1$ judges. 
Note here $v$ is defined as a three-value variable.
As we have informally discussed at the beginning of the section, a judge's protocol $\prot$ satisfies functionality, if
the system generated by $\prot$ satisfies the formula 
$AF(v=\fm^{2n+1}(d_0,d_1,\dots,d_{2n}))$, which is essentially a \emph{liveness} requirement. 
In our actual verification in MCMAS, we release our condition to the formula
$\bigwedge_{i\in\set{1,\dots n}} AF(\knows_i(v=1)\vee\knows_i(v=0))$, provided
that the protocol ensures that $v$ always gets the correct majority result.
The anonymity requirements to be discussed in the following paragraph are in general based on the fulfilment of 
functionality of a protocol. Our 
epistemic accessability relation $\sim_i$ is defined as $s\sim_it$ iff $\Obs_i(s)=\Obs_i(t)$,
where the observation function $\Obs_i(s)$ gives judge $i$ only the values of $v$ and $d_i$ at state $s$.

The definition of anonymity is more elaborated for the judges protocols, since sometimes 
it is impossible to prevent a judge from deducing the other judges' decisions by knowing the final verdict 
together with recalling his own decision, as we have already discussed above in the case of three judges.
Formally, we define \emph{conditional anonymity} in the form of 
\[AG(\phi_{i,j}\Rightarrow(\neg\knows_i(d_j=1)\wedge\neg\knows_i(d_j=0)))\] 
for all judges $i\neq j$, i.e., judge $i$ does not know judge $j$'s decision conditional to 
the formula $\phi_{i,j}$. In our protocol analyses in Section~\ref{sec:analysis},
we derive particular conditional anonymity requirements to serve in 
each different scenario. The following paragraphs present the strongest notions of 
anonymity that are not to be applied in the protocol analyses for more than three judges in Section~\ref{sec:analysis}. 
However, we believe they are of theoretical importance to be connected with other
definitions of anonymity in the literature (such as~\cite{HO03}). 

\paragraph{Perfect individual anonymity.}
Here we present an anonymity definition which requires that every judge $J_i$ is not allowed to deduce the decisions of 
every other judge $J_j$ in a run, if $J_j$'s decisions as `1' and `0' are both compatible with the final 
verdict $v$ as well as $i$'s local decision $d_i$. Note that this notion is essentially 
what we have presented as compatibility based anonymity at the beginning of the section. 
Formally, a protocol $\prot$ satisfies \emph{perfect individual anonymity}, 
if the system generated by $\prot$ satisfies that for all judges $i,j\in\set{0,\dots, 2n}$ with $i\neq j$,
if the value of $d_j$ cannot be derived from $d_i$ and $v$, then it cannot be deduced by $J_i$ at any
time during a protocol execution, i.e., 
\[    \bigwedge_{i,j\in\set{0,1,\dots,2n}}AG(\comp(i,j,v)\Rightarrow(\neg\knows_i (d_j=0)\wedge\neg\knows_i (d_j=1)))\]
where $\comp(d_i,d_j,v)$ denotes the compatibility between the decisions $d_i$, $d_j$ and the final verdict $v$,
which can be formally defined as that there exist boolean values $v_0,v_0',v_1,v_1'\dots, v_{2n},v_{2n}'\in\set{0,1}$ satisfying 
$v_i=v_i'=d_i$, $v_j=\neg v_j'=d_j$ and $\fm^{2n+1}(v_0,v_1,\dots,v_{2n})=\fm^{2n+1}(v_0',v_1',\dots,v_{2n}')=v$.
In our verification, this formula is usually split
into separate subformulas for each pair of judges. For example, in the case of three judges, we specify
$AG((v=d_0)\Rightarrow(\neg\knows_0(d_1=0)\wedge\neg\knows_0(d_1=1)))$ for judges $J_0$ and $J_1$,
and the other five formulas (by other ways of taking distinct $i,j$ out of $\set{0,1,2}$) 
can be specified in a similar way.

Equivalently, this specification can also be understood as 
if both $0$ and $1$ are possible for $d_j$ from the values of $d_i$ and $v$, then
both $0$ and $1$ are deemed possible by $J_i$ throughout the protocol execution. As the \emph{possibility modality} $P_i$ is 
defined as $P_i\varphi$ iff $\neg\knows_i(\neg\varphi)$, we can rewrite the condition in terms of possibility
similar to what is defined by Halpern and O'Neill~\cite{HO03}. For example, the above can also be
restated as
\[    \bigwedge_{i,j\in\set{0,1,\dots,2n},b\in\set{0,1}}AG(\comp(i,j,v)\Rightarrow P_i(d_j=b)). \]

\paragraph{Total anonymity.}
It is also possible to define an even stronger notion of anonymity. Let a \emph{decision profile} of size $n$ be a member of the set 
$\set{0, 1}^n$, and write $d(i)$ for the $i$-th member of a decision profile $d$ for $0\leq i\leq n-1$. 
Intuitively, a decision profile is a vector consisting of all judges' decisions. We overload the equivalence operator `$=$' 
by defining the equality $\List{d_0,d_1,\dots d_{n-1}}=d$ for decision profile $d$ of size $n$ holds iff $d_i=d(i)$ for all $i$,
i.e., the judges decisions $d_0, d_1,\dots d_n$ as a vector is equivalent to the decision profile $d$.
A protocol $\prot$ satisfies \emph{total anonymity}, if the following is satisfied,
\[
\bigwedge_{i\in\set{0,1,\dots,2n},d\in\set{1, 0}^{2n+1}}AG((d(i)=d_i\wedge\fm^{2n+1}(d(0),d(1),\dots,d(2n+1))=v)\Rightarrow P_i(\List{d_0,d_1,\dots d_{2n}}=d)). 
\]
That is, every judge cannot rule out every possible combination of decisions that is compatible with his own decision and the final verdict.
It is obvious that total anonymity and perfect individual anonymity are the same in the case of three judges,
but total anonymity is strictly stronger than perfect individual anonymity when there are more than three judges.
This notion can be shown as a special case of total anonymity of Halpern and O'Neill~\cite{HO03}.

We show that \emph{total anonymity} is strictly stronger than \emph{perfect individual anonymity} when there are five or more judges.
Suppose there is a protocol $P'$ for five judges where the decisions of $J_0$, $J_1$, $J_2$, $J_3$, $J_4$
are $d_0, d_1, d_2, d_3, d_4$, respectively. Assuming that the protocol is almost perfect, in the sense that
$J_i$ knows only the final verdict and his own decision throughout the protocol executions for all $i$ with $i\in\set{1,2,3,4}$,
but in the end $J_0$ knows in addition whether or not there are at least three other judges have voted for guilty. Protocol $P'$
satisfies \emph{perfect individual anonymity}, since for all $i,j\in\set{1,2,3,4,5}$, $J_i$ considers both $d_j=1$ and $d_j=0$ 
possible if $i\neq j$. However, $P'$ does not satisfy \emph{total anonymity}, because if the final verdict is guilty, $d_0=1$, and $J_0$ knows
there are less than three other judges who have voted for guilty, $\List{1,1,1,1,1}$ is not a deicision profile 
compatible with $J_0$'s observation in this run.

\section{Automatic Analysis in MCMAS}
\label{sec:analysis}

We have modelled and checked the above two solutions in MCMAS~\cite{LQR09},
which is a symbolic model checker supporting specifications in an extension of
CTL (computational tree logic) with epistemic modalities, including the modality $\knows$.
All anonymity properties we are interested in
have the form of conditional anonymity $AG(\phi\Rightarrow\neg\knows_i\varphi)$
(see Section~\ref{ssec:conditionalanonymity}), where $\phi$ typically represents the final outcome
of a protocol and/or what the individual decision of $J_i$ and
$\knows_i\varphi$ represents the knowledge of $J_i$ ($\knows_i\varphi$ means ``agent $i$ knows $\varphi$").

The input language ISPL (Interpreted Systems Programming Language) of MCMAS
supports modular representation of agent-based systems.
An ISPL agent is described by giving the agents' possible local states,
their actions, protocols, and local evolution functions.
An ISPL file also defines the initial states, fairness constraints, and properties to be checked.
The interpretation for the propositional atoms used in the properties can also be given.
The semantics of an ISPL file is an interpreted system,
upon which interesting properties are defined as well.
(Details about MCMAS and ISPL can be found in~\cite{LQR09}.)
How we model the two solutions to the judges problem in ISPL is out of the scope of the current paper.
Instead, we focus on the anonymity properties and their model checking results in MCMAS.
Functionality properties can be simply checked by
comparing all individual judge's decision and the final verdict of the protocol.

\paragraph{Analysis of anonymity properties in the centralised solution.}

Due to the different roles an agent can play, either the leader or not,
we define a conditional anonymity property for judges in a different way.
For any $J_i$ ($i\neq 0$) who is not the leader,
it should be the case that he does not know anything about any other judge's decision.
This is formalised as a logic formula $AG( \neg \knows_i (d_j=1)\wedge\neg\knows_i(d_j=0))$ with $i\neq j$.
Ideally, this formula should also hold for $J_0$ as well, who plays the lead role.
However, this is not the case, as $J_0$ collects the bits
$d^{\land}_{2i-1}$, $d^{\lor}_{2i-1}$, $d^{\land}_{2i}$, $d^{\lor}_{2i}$
from every pair of judges $(J_{2i-1},J_{2i})$ ($1\leq i\leq n$).
If both $J_{2i-1}$ and $J_{2i}$ have made the same decision $d_{2i-1}=d_{2i}$,
the leader $J_0$ would find it out by simply checking the values of $d_{2i-1}\land d_{2i-1}$ and $d_{2i-1}\lor d_{2i-1}$.\footnote{
Both $d_{2i-1}\land d_{2i-1}$ and $d_{2i-1}\lor d_{2i-1}$ are true, if $d_{2i-1}=d_{2i}=1$;
both $d_{2i-1}\land d_{2i-1}$ and $d_{2i-1}\lor d_{2i-1}$ are false, if $d_{2i-1}=d_{2i}=0$.}
Hence, for $J_0$, the anonymity property is formalised
as a logic formula 
\[AG((d_{2i-1}\not=d_{2i})\Rightarrow (\neg \knows_0 (d_{2i-1}=0) \land \neg \knows_0 (d_{2i-1}=1)\land\neg \knows_0 (d_{2i}=0) \land \neg \knows_0 (d_{2i}=1) ))\]
by excluding the above situations from the premise.

In case of a protocol with more than or equal to $5$ judges,
both properties are checked to hold in MCMAS.
A protocol with three judges is a special case.
First, it is not necessary for the leader $J_0$ to obtain all
$d^{\land}_{1}$, $d^{\lor}_{1}$, $d^{\land}_{2}$, $d^{\lor}_{2}$ as seen in Section~\ref{ssec:mm-solution}.
Second, if one of the judge's decision $d_i$ is `guilty' (`innocent') and
the final verdict $v$ is `innocent' (`guilty'), then this judge can find out that
the other two judges have voted for `innocent' (`guilty').
Hence, we need a different formalisation
\[\bigwedge_{i\neq j}\quad AG((v=d_i) \Rightarrow (\neg \knows_i (d_j=0)\wedge\neg\knows_i(d_j=1)) ).\] 

\paragraph{Analysis of anonymity properties in the DCP-based solution.}

In the DCP-based solution, the final verdict $v$ is the number of votes for `guilty'.
As discussed in Section~\ref{ssec:conditionalanonymity},
the definition of anonymity has to take care of the possibility that 
a judge can deduce the other judges' decisions from the final verdict together with his own decision.
For example, if the final verdict is $2n+1$ (or $0$), then
it should be the case that every judge has voted for `guilty' (`innocent').
Another situation is that if the final verdict is $2n$ (or $1$), 
and one judge's decision is `innocent' (or `guilty'),
then this judge can know that every other judge's decision is `guilty' (or `innocent').\footnote{
It is also possible that a group of judges cooperate together
to find out the rest judges' decisions.}
Hence, for the DCP-based protocols anonymity is formalised as
\[\bigwedge_{i\neq j}\quad AG( ((1< v< 2n) \lor 
(v=1 \land d_i=0) \lor (v=2n\land d_i=1))\Rightarrow (\neg \knows_i (d_j=0)\wedge\neg\knows_i(d_j=1)) ).\]

\paragraph{Summary of verification results in MCMAS.}
All aforementioned conditional anonymity properties have been checked successfully on instances of the two solutions
with three or five judges, respectively.
Table~\ref{tab:results} summarizes the statics in MCMAS (version 0.9.8.1).
The large increase in states and BDD memory consumption in the case of
DCP-based protocols is due to the use of arithmetic operations.
Extending the models for more judges is an interesting exercise in MCMAS,
but it is not the focus of the current paper.

\begin{table}
\label{tab:results}
\begin{tabular}{|l||r|r||r|r|}
\hline
 &\multicolumn{2}{c||}{The centralised solution}&\multicolumn{2}{c|}{The DCP-based solution}\\
\cline{2-5}
 & reachable states & BDD memory (MB) & reachable states & BDD memory (MB)\\
\hline\hline
3 judges & 184 & 4.90 &  95,972 & 5.70 \\
\hline
5 judges & 5,568 & 5.10 & 1.43$\times 10^8$ &  55.00 \\
\hline
\end{tabular}
\end{table}

\section{Discussion and Future Work}
\label{sec:conclusion}

In the current paper, we have presented two solutions to the judges problem
to compute a majority function securely.
One solution is based on the original proposal by McIver and Morgan~\cite{MM09} using oblivious transfer,
the other is an extension of the DCP~\cite{chaum_88} for compute the sum of the judges' decisions.
Both are imperfect in the sense that judges are not unconditionally anonymous,
some of judges can obtain more information than their own decisions.
This has been captured by our notion of conditional anonymity and confirmed by
the automatic analysis in a model checker.

In the literature, the question about secure multi-party computation was originally suggested by Yao~\cite{yao_82},
with which he presented the \emph{millionaires problem}. The problem can be stated as that
two millionaires want to find out who is richer without revealing the precise amount of their wealth.
Yao proposed a solution allowing the two millionaires to satisfy their curiosity while respecting their privacy.
Further generalisations to Yao's problem are called multi-party computation (MPC) protocols,
where a number of parties $p_1, p_2, \ldots, p_n$, each of which has a private data
respectively $x_1, x_2, \ldots, x_n$,
want to compute the value of a public function $F(x_1, x_2, ..., x_n)$.
An MPC protocol is considered secure if no party $P_i$ can learn more than the description of the public function,
the final result of the calculation and his own $x_i$.
The judges problem is just a special MPC protocol for computing a majority.
The security of such kind of protocols can be either computational or unconditional.
In most part of this paper we focus on the latter case.
It is also of interest to derive computational solutions,
as communicated with Radomirovi{\'c}~\cite{sasa_09}.
For instance, Brandt gives~\cite{Bra05} a general solution for securely computing disjunction and
maximum for both active and passive attackers, base on El-Gamal encryption.
Chor and Kushilevitz~\cite{CK93} study the problem of computing modular sum when the inputs are distributed.
Their solution is \emph{$t$-privately},
meaning that no coalition of size at most $t$ can infer any additional information.
A generalisation has been made by Beimel, Nissim and Omri~\cite{BNO08} recently.
It would be interesting to see if these schemes can be used also for computing the majority function.
In the appendix, we present one possible computational solution based on the anonymous veto networks~\cite{hao_09}.
This solution does not take efficiency into account,
while lower bounds on message complexity are given in~\cite{CK93,BNO08}.
How to achieve a most efficient solution to securely compute a majority function is one of our future work.
More importantly, having a formal correctness argument,
for instance with the support of a theorem prover, is another future work.

Recently, the population protocol model~\cite{AR07} has emerged as an elegant computation paradigm
for describing mobile ad hoc networks, consisting of multiple mobile nodes which interact with
each other to carry out a computation.
One essential property of population protocols is that with respect
to all possible initial configurations all nodes must eventually
converge to the correct output values (or configurations).
To guarantee that such kind of properties can be achieved, the interactions of nodes in population
protocols are subject to a strong fairness ---
if one action is enabled in one configuration, then
this action must be taken infinitely often in such a configuration.
The fairness constraint is imposed on the scheduler to ensure that the protocol makes progress.
In population protocols, the required fairness condition will make 
the system behave nicely eventually, although it can behave
arbitrarily for an arbitrarily long period~\cite{AR07}.
Delporte-Gallet {\em et al.}~\cite{delporte_07} consider private computations in the population protocol model.
The requirement is to compute a predicate without revealing any input to a curious adversary.
They show that any computable predicate, including the majority function,
can be made private through an obfuscation process.
Thus, it is possible to achieve a solution to the judges problem within their framework.
After that, we can formally model check the solution in the tool PAT~\cite{DLSP09},
which is dedicated to deal with fairness conditions for population protocols.
However, as discussed above the population protocol can only guarantee a majority eventually computed.
But for agents (judges) in the protocols, they have no idea of when this is successfully computed.
Whether this is a desirable solution of the judges problem is still under discussion.
Moreover, we have only considered \emph{curious but honest} judges.
It is interesting to extend the available solutions to
take active adversaries and/or coalition of dishonest judges into account, e.g., following~\cite{Bra05,CK93}.

\paragraph{Acknowledgement.}
We are grateful to Carroll Morgan for sharing his three-judges protocol with us and useful discussions.
We thank Sas\v{a} Radomirovi{\'c} for many discussions on secure multi-party computation
and Hongyang Qu for helping us with MCMAS.
We also thank the anonymous referees for their valuable comments.

\bibliographystyle{plain}
\bibliography{judges}


\appendix


\newcommand\Rin{\in_{R}}
\newcommand\mod{\texttt{mod}}

\section{A Computationally Anonymous Majority Function Protocol}

The idea of this protocol is partially
due to Sa\v{s}a Radomirovi\'c. 
The functionality of the protocol relies on the Anonymous Veto Network~\cite{hao_09}.
The protocol assumes a finite cyclic group $G$ of prime order $q$, and the judges of $J_0\dots J_{2n}$ 
agree on a generator $g$ of $G$. The values of $g$ and $q$ are larger enough ($\gg n$).
The protocol consists of three steps of computations by means of broadcast, and
there are no private channels required. The first step (round) sets up a nonce $g^{N_i}$ for each 
judge $i$ satisfying $\sum N_i = 0_{ \mid q}$. 
The second step (which actually takes $n$ rounds) pre-computes the $n+1$ majority values 
in an encrypted form of $g^{n+1}$, $g^{n+2}$, $\dots g^{2n+1}$, which are secretly shuffled 
so that no judge knows which one is which. In the last step every judge announces
his vote in a secure way, so that the final verdict will be known by every judge by 
examining whether the final result is within the set of the pre-computed values 
from step two. The protocol can be formally stated as follows.
\paragraph{Step 1} 
Every judge $J_i$ publishes $g^{x_i}$ and a zero knowledge proof of $x_i$. After that,
every judge is able to compute 
\[g^{y_i}=\prod_{j=0}^{i-1}g^{x_j}/\prod_{j=i+1}^{2n}g^{x_j} \]
Now for each judge $i$, he has $g^{N_i}=g^{x_iy_i}$ satisfying $\sum N_i = 0$ ($\mod\ q$), which is equivalently
\[\prod_{i=0}^{2n}g^{N_i}=1\]
\paragraph{Step 2}
Let $M=\set{n+1,n+2,\dots, 2n+1}$ be the set of majority values. Every judge $J_i$ generates a
random permutation $p_i:M\rightarrow M$. Then judge 1 computes $m_{k,0} = g^{k}$ for all
$k\in M$. Subsequently, in the precise order from $J_0$ to $J_{2n}$, 
$J_i$ announces the sequence $\List{m_{n+1,i}, m_{n+2,i}, \dots m_{2n+1,i}}$,
where $m_{p_i(k),i}= (m_{k,i-1})^{x_i}$ for each judge $J_i$. Write $M'$ for the
final set $\set{m_{k,2n+1}}_{k\in M} = \set{g^{kx_1x_2\dots x_{2n+1}}}_{k\in M}$. Intuitively,
the members in $M'$ are randomly shuffled such that no judge knows what each individual value in $M'$
originally corresponds to in $M$.
\paragraph{Step 3}
Each judge $i$ decides his vote $v_i\in\set{0,1}$, and publishes $z_{i,1} = g^{(N_i+v_i)x_i}$.
This requires additional $2n$ rounds, and for each round $r\in\set{2,3,\dots,2n+1}$, Judge $J_i$ takes $z_{i\ominus 1,r-1}$
where $i\ominus 1=i-1$ if $i\geq 1$ and $0\ominus 1 = 2n$, and publishes $z_{i,r}$ as $(z_{i\ominus 1,r-1})^{x_i}$.
Finally we have the results as a sequence 
$\List{z_{2n+1,2n+1},z_{1,2n+1},z_{2,2n+1},\dots z_{2n,2n+1}}$ such that 
$z_{i,2n+1} = g^{(N_{i\ominus 1}+v_{i\ominus 1})x_1x_2\dots x_{2n+1}}$.
Every judge then can check if 
\[\prod_{i=0}^{2n}z_{i,2n+1}\in M'\]
If yes then the final verdict is `yes' (guilty), otherwise the final verdicit is `no' (innocent).

\end{document}